\documentclass[10pt]{article}
\pdfoutput=1
\usepackage{amsmath}
\usepackage{amssymb}

\usepackage{graphicx}

\usepackage{cite}
\usepackage[usenames,dvipsnames]{color}

\usepackage[labelfont=bf,labelsep=period,justification=raggedright]{caption}

\usepackage[table]{xcolor}
\definecolor{lightgray}{gray}{0.9}

\makeatletter
\renewcommand{\@biblabel}[1]{\quad#1.}
\makeatother

\date{}

\begin{document}

\begin{flushleft}
{\Large
\textbf{Predicting chemical environments of bacteria from receptor signaling}
}
\\
Diana Clausznitzer$^{1,2,3,4,\dagger}$, 
Gabriele Micali$^{1,2,\dagger}$,
Silke Neumann$^{5}$,
Victor Sourjik$^{5,6}$ and 
Robert G. Endres$^{1,2,\ast}$
\\
\bf{1} Department of Life Sciences, Imperial College, SW7 2AZ London, UK
\\
\bf{2} Centre for Integrative Systems Biology and Bioinformatics, Imperial College, SW7 2AZ, London, UK
\\
\bf{3} BioQuant, Heidelberg University, 69120 Heidelberg, Germany
\\
\bf{4} Institute for Medical Informatics and Biometry, Technische Universit\"at Dresden, 01307 Dresden, Germany
\\
\bf{5} Centre of Molecular Biology, Heidelberg University, DKFZ-ZMBH Alliance, 69120 Heidelberg, Germany
\\
\bf{6} Max Planck Institute for Terrestrial Microbiology, D-35043 Marburg, Germany
\\
$\dagger$ Both authors contributed equally to this work
\\
$\ast$ E-mail: r.endres@imperial.ac.uk
\end{flushleft}

\section*{Abstract}
Sensory systems have evolved to respond to input stimuli of certain statistical properties, and to reliably transmit this information through biochemical pathways. Hence, for an experimentally well-characterized sensory system, one ought to be able to extract valuable information about the statistics of the stimuli. 
Based on dose-response curves from {\it in vivo} fluorescence resonance energy transfer (FRET) experiments of the bacterial chemotaxis sensory system, we predict the chemical gradients chemotactic \textit{Escherichia coli} cells typically encounter in their natural environment. To predict average gradients cells experience, we revaluate the phenomenological Weber's law and its generalizations to the Weber-Fechner law and fold-change detection. To obtain full distributions of gradients we use information theory and simulations, considering limitations of information transmission from both cell-external and internal noise. 
We identify broad distributions of exponential gradients, which lead to log-normal stimuli and maximal drift velocity. 
Our results thus provide a first step towards deciphering the chemical nature of complex, experimentally inaccessible cellular microenvironments, such as the human intestine.

\section*{Introduction}

Bacteria live in complex chemical microenvironments with the spatio-temporal chemical concentration profiles shaped by nutrient supply, as well as nutrient depletion by surrounding bacteria of the same or other species (Fig.~1).
From studies of the composition of commensal bacteria in our intestines, it is emerging that these bacterial communities have important roles for breaking down undigested food, and for increasing resistance against colonization by pathogens~\cite{NatInNat05}.
Although of physiological importance, we currently do not have a good understanding of what chemical gradients microorganisms in the gut typically encounter.
One way forward is to exploit our accumulated knowledge about their sensory systems, evolved to detect typical stimuli relevant to the organism, e.g. the concentration of a particular nutrient. Therefore, knowing the design and functioning of sensory systems we ought to be able to predict typical stimuli such as concentration gradients.

The best-characterized bacterium is {\it Escherichia coli}, inhabiting the gastrointestinal tract of humans and animals. This bacterium grows in the mucus layer secreted by the intestinal epithelium~\cite{PouLanKrog94,Gau07}.
The chemotaxis sensory system enables these bacteria to detect and migrate in chemical gradients of nutrients and toxins. 
Unlike many sensory systems, it is relatively simple with only a few components and well understood at the molecular level \cite{Berg00,FalHaz01,Sou04,WadArm04,BakWolStock06}.
This and other sensory systems share common design principles including Weber's law, the Weber-Fechner law and fold-change detection (FCD)~\cite{Lau87,Lau89, OlsFie00,JohHsiYosh02, Deh03, DunnRieke06, Kojadinovic_2013} (see Text S1 and Fig. S1A for an introduction). Weber's law predicts that the minimum change in stimulus perceived against a background stimulus increases proportionally to the background stimulus, thus implementing efficient contrast coding~\cite{MesOrdAdl73,KalTuWu09,OlsFie00}. The integrated version of Weber's law is the Weber-Fechner law, stressing the internal representation of stimuli~\cite{JohHsiYosh02}.
FCD is a generalization of Weber's law to large stimuli and dynamics~\cite{ShovGoeAlon10} that postulates the invariance of responses to time-dependent stimuli when scaled up or down by a constant factor~\cite{Kojadinovic_2013, LazAhmShi2011}. 
While these laws relate to average gradients~\cite{LazAhmShi2011, VlaLovSou08}, they do not quantify the complexity of the chemical environment in terms of distributions of concentrations and gradients, or provide information on their functional importance in terms of cell-swimming behavior.

To fully characterize the chemical environment of chemotactic bacteria in terms of distributions, we expect a matching between the statistical properties of typical stimuli and the corresponding cellular responses assuming certain optimization criteria. 
For instance, taking into account cellular limitations of signal processing due to noise and limited energy resources, a reasonable assumption is that cells translate a stimulus (input) into a cellular response (output) in a way that maximizes the transmission of information~\cite{Lau81, DetRamShr00, TkaCal07a,Mehta_MSB_2009}.
An appropriate measure for the information contained in the output about the input is Shannon's mutual information~(see Text S1 and Fig. S1B for an introduction to information theory)~\cite{BialekBook}, which effectively describes the correlations between the inputs and outputs in bits, with a fundamental link to thermodynamics \cite{Toyabe10}. It is furthermore unclear how the above mentioned phenomenological laws follow from first principles, such as information theory.
%

Here, we demonstrate the ability to predict distributions of typical gradients of chemoattractant aspartate {\it E. coli} likely encounters in its physiological microenvironment.
For this purpose we use quantitative dose-response curves from {\it in vivo} fluorescence resonance energy transfer (FRET) measurements of the chemotactic sensory pathway, information theory, as well as measurements and simulations of swimming bacteria. 
Using our multidisciplinary approach, we show that maximizing mutual information implies maximizing the alignment of cell trajectories with the gradient (chemotactic index), and hence the drift velocity of cells swimming up the gradient. We thus go significantly beyond Weber's law, Weber-Fechner law and fold-change detection, which only predict average chemical gradients. Our approach is a first step towards the prediction of typical chemical environments of bacteria, shaped by cellular communities and dynamic turnover of nutrients. These environments are difficult to determine experimentally, but are biomedically highly important for gut metabolism, tissue homeostasis, and disease development in the gastrointestinal tract \cite{Sommer_13,Kamada_13}.

\section*{Results}

\subsection*{Prior knowledge of receptor signaling}

{\it E. coli} chemotaxis is well known for its high sensitivity to detect minute changes in chemical concentration (attractant and repellents) due to cooperative receptor complexes \cite{DukeBray99}. Furthermore, cells adapt precisely to persistent chemical stimulation due to covalent receptor modification (methylation and demethylation) \cite{BarLei97}. As a result, cells can chemotax in a wide range of chemical concentrations. The final pathway output are the flagellated rotary motors, which lead to runs (swimming) or tumbling (random reorientation) \cite{BergToday00}, and hence biased random walk in the gradient.

The Monod-Wyman-Changeux (MWC) model was developed to describe the chemoreceptor signaling activity~\cite{EndSouWin2008,SouBerg04,MelTu05,KeyEndSko06,EndWin06}.
The average signaling activity of a receptor complex composed of $N$ receptors, which can switch between an {\it on} and an {\it off} state, is given by
\begin{equation}
 A = \frac{1}{1+e^{F(c, m)}},\label{eq:mwc}
\end{equation}
where the free-energy difference $F(c, m)$ between the {\it on} and {\it off} state is
\begin{align}
F(c, m) = & N \bigg[ \epsilon(m) + \nu_a \ln\left( \frac{ 1+c/K_a^{\text{off}} }{ 1+c/K_a^{\text{on}} } \right)  + \nu_s \ln\left( \frac{ 1+c/K_s^{\text{off}} }{ 1+c/K_s^{\text{on}} } \right) \bigg]
\label{eq:DF}
\end{align}
with $\epsilon(m) \simeq 1 - \frac{1}{2} m $~\cite{EndSouWin2008}. (See {\it Materials and Methods} for parameters value). 
The free-energy difference is a function of the ligand concentration $c$ and the average modification level $m$ of a receptor in the complex.
Here, we consider two receptor types mixed in a complex, Tar (indicated by index $a$) with fraction $\nu_a$ of receptors in the complex, and Tsr (indicated by index $s$) with fraction $\nu_s$ of receptors. Receptors are sensitive to attractant with dissociation constants $K^\text{on}_{a/s}$ and $K^\text{off}_{a/s}$ in the {\it on} and {\it off} state, respectively. Wild-type cells have a receptor ratio of $\nu_a$:$\nu_s$=1:1.4~\cite{KeyEndSko06}, whereas Tar-only cells have $\nu_a$=1 and $\nu_s$=0. This model quantitatively describes the signaling activity of non-adapting cells as measured by FRET \cite{KeyEndSko06}.

In adapting cells, we additionally need to describe the process of receptor modification dynamically. 
In this model, we describe the dynamics by the following ordinary differential equation~\cite{ClauOleEnd10}
\begin{equation}
 \frac{dm}{dt}=g_R (1-A) - g_B A^3,\label{eq:preciseadaptation}
\end{equation}
with $g_R$ and $g_B$ the methylation and demethylation rate constants from enzymes CheR and phosphorylated CheB, respectively. According to this model, receptors are methylated when the complex is inactive and demethylated when the complex is active, leading to precise adaptation~\cite{BarLei97,EndWin06} (see {\it Materials and Methods} for parameter values and Fig. S2). The model in Eq. \ref{eq:preciseadaptation} is able to describe a wide range of time-course data from FRET but other dependencies of the rates on activity $A$ may also work \cite{Shimizu10MSB}.

Using multiple measurements of the response to the same stimulus from FRET \cite{SouBerg02a}, we characterize the input-output relationship of receptor signaling, including noise from the experiments and the sensory system, respectively, for an information-theoretic approach.
Specifically, we use the fast response of chemoreceptors and intracellular signaling to the 
chemoattractant $\alpha$-methyl-DL-aspartate (MeAsp), a non-metabolizable analogue of aspartate, to characterize the initial steps in the chemotaxis sensory system. Two different data sets were used in this paper:
Firstly, dose-response curves for adapting wild-type cells (WT 1) expressing both Tar and Tsr receptors, shown in Fig.~2A and B~\cite{ClauOleEnd10}. This data is ideal to investigate Weber's law, as sensory adaptation adjusts the sensory system to the current background attractant concentration, and step changes in concentration probe the instantaneous intracellular signaling response.
Secondly, we use dose-response curves shown in Fig.~3A~\cite{EndSouWin2008} of mutants expressing only the Tar chemoreceptor which are either adapting (WT 2) or genetically engineered to have the amino acids glutamate (E) or glutamine (Q) at the four receptor modification sites for methylation and demethylation. Specifically, E is an unmethylated modification site, while Q resembles a methylated modification site, allowing us to mimic the various methylated states receptors can assume. Hence, these engineered bacterial strains are more tractable experimentally, enabling us to study the receptor response in defined adaptation states.

\subsection*{Fold-change detection and Weber's law} 

To obtain information about the preferred average gradients of {\it E. coli} cells, we first critically reevaluate fold-change detection (FCD) and Weber's law in {\it E. coli} chemotaxis using our data. FCD is characterized by the invariance of the response to scaling up or down the ligand concentration. This means that when the background ($c_0$) and difference ($\Delta c$) concentrations are scaled by the same factor, i.e. $c_0 \rightarrow \gamma c_0$ and $\Delta c \rightarrow \gamma \Delta c$, the response amplitude remains the same. Expressing the change in free-energy difference upon stimulation of initially adapted receptors as 
\begin{eqnarray}
 \Delta F 
    =  N \left[ \nu_a \ln\left( \frac{ 1+\frac{\Delta c}{K_a^{\text{off}}+c_0} }{ 1+\frac{\Delta c}{K_a^{\text{on}} + c_0} } \right) + \nu_s \ln\left( \frac{ 1+\frac{\Delta c}{K_s^{\text{off}}+c_0} }{ 1+\frac{\Delta c}{K_s^{\text{on}} + c_0} } \right) \right],
\label{eq:DeltaF}
\end{eqnarray}
the ratios in the logarithm can be re-written 
\begin{equation}
\frac{\Delta c }{ K_{a,s}^{\text{off,on}}+c_0 }= \frac{\Delta c }{ c_0} \frac{1}{ 1 + K_{a,s}^{\text{off,on}} / c_0},
\end{equation}
which is only a function of the fraction $\Delta c / c_0$ in two concentration regimes: ({\it i}) $K_a^{\text{off}} \ll c_0 \ll K_a^{\text{on}}$ and ({\it ii}) for $K_s^{\text{off}} \ll c_0 \ll K_s^{\text{on}}$. Hence, the change in free-energy difference $\Delta F$, as well as the change in receptor-complex activity $\Delta A$, are only a function of the fractional concentration change for background concentrations between the dissociation constants of  each receptor type, i.e. when receptor are most responsive. The change in free-energy difference in these regimes is
\begin{equation}
\Delta  F \approx \begin{cases} %
     N\nu_a \ln\left(1+\frac{\Delta c}{c_0}\right) & \mbox{if } K_a^{\text{off}} \ll c_0 \ll K_a^{\text{on}} \text{ (regime I)}\\
     N\nu_s \ln\left(1+\frac{\Delta c}{c_0}\right) & \mbox{if } K_s^{\text{off}} \ll c_0 \ll K_s^{\text{on}} \text{ (regime II)}\\
\end{cases}\label{eq:FCD}
\end{equation}
and $\Delta F\approx 0$ otherwise. Therefore, there are two concentration regimes corresponding to the responsive ranges of the two receptor types, Tar and Tsr, for which fold-change detection is expected. Indeed, these two regimes of FCD are consistent with recent experiments for small periodic stimuli \cite{LazAhmShi2011}. Based on our ligand dissociation constants from the {\it Materials and Methods} section, our regime I corresponds to their first FCD regime, the transition from our regime I to regime II coincides with their second FCD regime. Finally, our regime II corresponds to their osmotic stress regime. Our interpretation is supported by the low response amplitude of their second FCD regime (see also Fig. 3B in \cite{EndWin06}).

To investigate FCD even for large stimuli, we use FRET time courses and dose-response curves, as well as the dynamic MWC model for adapting wild-type cells (WT 1). Fig.~2A and B define the adapted activity, adaptation time and change in activity $\Delta A$. Specifically, adaptation time is defined here as the duration from the onset of the concentration change to the time at which the response is restored to half the pre-stimulus level. Fig.~2C shows that FCD approximately holds for small response amplitudes $\Delta A$ since all the dose-response curves approximately collapse onto a single curve when plotted as a function of the fractional concentration change. 
However, FCD requires that also adaptation time only depends on the fractional change in concentration to yield an invariant temporal response. From our model, we expect FCD to extend to adaptation time as the adaptation dynamics in Eq.~\ref{eq:preciseadaptation} only depend on the receptor complex activity. 
Fig.~2D shows that similar to the response amplitudes, small fractional changes yield similar adaptation times (cf.~\cite{LazAhmShi2011}) with significant deviations at large fractional changes. 
These deviations might be due to the fact that large concentration changes $\Delta c$ exceed regime I. The breakdown of FCD with respect to adaptation time may additionally reflect the previous observation in single cells that the precision of adaptation is robust, but not the adaptation time~\cite{AlonSurLeib99}.

Weber's law is a special case of FCD for small fractional changes. Using the FRET dose-response curves and the dynamic MWC model for adapting wild-type cells (WT 1) we extract the threshold stimulus for different background concentrations (Fig.~2B): we define the threshold stimulus $\Delta c_t$ to be the concentration increment which produces the noticeable activity response $\Delta A$ relative to the adapted level $A^*$. 
Experimental dose-response curves and our modeling indeed confirm the validity of Weber's law for the chemotactic sensory system (Fig.~2E).
Note the precise value of $\Delta A$ is not essential in obtaining Weber's law (Fig. 2E, inset) but can also be estimated from theory (see Text S1). To obtain an analytical formula for Weber's law for given $\Delta A$, we Taylor-expand Eq.~\ref{eq:mwc} for the receptor-complex activity up to linear order about the steady-state activity $A^*$, leading to $\Delta A \sim \Delta c_t/c_0$ (see Text S1).
This explicitly shows that Weber's law $\Delta c_t \sim c_0$ is valid in both FCD regimes. 

\subsection*{Weber-Fechner law predicts chemotactic perception} 
 
The Weber-Fechner law predicts an internal representation of the ligand concentration, perception $R$, to follow $R\sim\ln c$~\cite{JohHsiYosh02}. This law results from the postulate that the threshold concentration for a particular background concentration corresponds to an increment $\Delta R$ in the internal representation of the concentration in the sensory system. This increment is a function of the fractional change in concentration $\Delta c / c$. Hence, integrating $\Delta R \sim \Delta c / c$ yields $R\sim\ln c$.

To derive the Weber-Fechner law we cannot integrate the activity response $\Delta A$ (Eq.~8 in Text S1) to obtain the perception with its logarithmic dependence (since factor $\partial A/ \partial  F$ depends implicitly on the attractant concentration). However, $\Delta F=k\cdot \Delta c / c$ (Eq. \ref{eq:DeltaF} for small $\Delta c / c_0$) can be integrated to result in the logarithmic perception $R$ given by the receptor complex free-energy difference
\begin{equation}
F \approx N \cdot \left[ \epsilon(m) + \nu_a \ln \left( \frac{c}{K^{\text{off}}_r} \right)\right],\label{eq:logDF}
\end{equation}
valid for concentrations $K^{\text{off}}_r \ll c \ll K^{\text{on}}_r$, with $r=a,s$ (cf. Eq.~\ref{eq:DF}). 
In the insets of Fig.~2C and D, we show the dose-response curves and adaptation times as function of a change in perception $\Delta R$. Clearly, the resulting data collapse holds well. This indicates that perception is a quantity relevant to the cell, thus favoring Weber-Fechner law over fold-change detection. Note the free-energy difference, Eq. \ref{eq:logDF}, was previously identified as the cause of logarithmic sensing \cite{KalTuWu09}.
As shown in Fig. S3A perception depends on the receptor-modification level $m$. At constant concentration, adaptation produces a perception equal to $F^*$, corresponding to adapted activity $A^*$. A sudden concentration change moves the perception along the logarithmic curve for a specific modification level with the perception deviating from adapted value $F^*$. 
Subsequent adaptation shifts the perception according to the $m$-dependent term in Eq. \ref{eq:logDF} so that $R$ returns to $F^*$ for the new concentration. 

\subsection*{Predicting typical gradients from Weber-Fechner law} 

To identify the gradients bacteria likely experience in their natural environment, we compare how moving bacteria perceive different gradients. We assume that spatial gradients which can be perceived over a wide range of concentrations, i.e. in which the perception neither diminishes due to adaptation nor saturates, may be the typical gradients the sensory system has evolved to detect. 
To gain intuition we consider a number of spatial gradients which arise from diffusion processes.
Free diffusion of a fixed number of ligand molecules deposited at a point  produces a Gaussian gradient~\cite{VlaLovSou08,BergInRandomWalks93}. Diffusion from a point source expelling molecules at a constant rate results in a hyperbolic gradient~\cite{BergInRandomWalks93}. Diffusion between points with constant rates of production and absorption produces a linear gradient~\cite{VlaLovSou08,HasPost07}. Finally, diffusion from a constant source and homogeneous degradation in the medium results in an exponential gradient~\cite{IbaBel08}.
Fig.~2F shows examples of these concentration gradients, where we have kept the value of the concentration the same at two points in space to make the gradients comparable.

We consider simple unidirectional one-dimensional swimming of bacteria with constant velocity $v_s$=20 $\mu$m~s$^{-1}$. This way, we can translate the spatial gradient into a temporal gradient, which is easier to analyze. 
To make progress, we approximate the free-energy difference by its logarithmic form Eq.~\ref{eq:logDF}, which is valid for the range of concentrations where Weber's law applies, e.g. $K^{\text{off}}_a \ll c \ll K^{\text{on}}_a$.
We find that an exponential gradient is perceived as constant as previously found by experiment~\cite{KalTuWu09} and computer simulation~\cite{VlaLovSou08} (Fig. S3B). In Supporting Text S1 we further demonstrate this for cells in exponential ramps in line with data \cite{BlockSegBerg83} (Fig. S3C,D). In contrast, the perception of the other gradients either attenuates by sensory adaptation (Gaussian and linear gradients), or increases, eventually saturating the response (hyperbolic gradient) along the swimming path (Fig.~S3B).
This may indicate that chemotactic {\it E. coli} typically encounters exponential gradients in its natural habitat, since its pathway has evolved to maintain perception in these. Although the Weber-Fechner law makes no predictions about the distribution of gradients, i.e. their statistics of occurrence, its focus on perception views changes in concentration as bearers of useful information for the cell. In the following we quantify this view of signaling.


\subsection*{Maximizing information transmission with noise}

Sensory systems convert inputs (here ligand concentration) into cellular outputs (here receptor activity) using noisy signal transduction pathways. Furthermore, inputs themselves carry uncertainty, and this input noise is transmitted through the pathway as well. Hence, to deal with the noise information transmission in a cell needs to be described in terms of probability distributions. Particularly useful is the mutual information, expressed as a function of the probability distributions of inputs and outputs, $p_c(c)$ and $p_A(A)$, respectively, the joint probability distribution $p(c, A)$, and the conditional probability distribution for output $A$ given input $c$, $p(A|c)$ (see Text S1 for details). 

Here, we assume that for a given input value $c$ the output of a single cell is distributed normally around a mean output value $\bar{A}(c)$ with variance $\sigma^2_T(c)$ due to transmitted input (cell-external) and output (cell-internal) noise. We further assume that the transmitted input ($\sigma^2_c$) and output ($ \sigma^2_A$) noise are independent. Hence, the output variance is given as the sum $\sigma^2_T = \sigma^2_A + (\partial \bar{A} / \partial c)^2 \sigma^2_c$.
The mutual information can be calculated analytically assuming that the noise in the output is small (an assumption we test later).
The mutual information is then given by (cf.~\cite{DetRamShr00})
\begin{equation}
 {\cal{I}}[c; A] = -\int \!\! dc \, p_c(c)\log_2\left[\sqrt{2\pi e [\sigma^2_A  (\partial \bar{A} / \partial c)^{-2}  + \sigma^2_c ]}p_c(c)\right],\label{eq:mutualinformation_gaussianchannel}
\end{equation}
i.e. the mutual information depends only on the distribution of inputs, the mean input-output relationship $\bar{A}(c)$ and the variances of input and output noise.
If we assume that the sensory system maximizes the mutual information between concentration inputs and cellular outputs, the relation between inputs and outputs, i.e. $\bar{A}(c)$, fulfils the following relationship (for derivation see {\it Materials and Methods}):
\begin{equation}
 \frac{\partial \bar{A}}{\partial c} \cdot \left[ 1 + \frac{\left( \frac{\partial \bar{A}}{\partial c}\right)^2  \sigma_{c}^2 }{\sigma_A^2 }\right] = Z \sigma_A \cdot p_c(c) ,\label{eq:optimalDR}
\end{equation}
where $Z$ is a constant given by the normalization of the input distribution
\begin{equation}
Z = \int\!\! dc \,\, \frac{1}{\sigma_A} \cdot \left( \frac{\partial \bar{A}}{\partial c}\right) \cdot\left[ 1 + \frac{\left( \frac{\partial \bar{A}}{\partial c}\right)^2  \sigma_{c}^2 }{\sigma_A^2 }\right].
\end{equation}
Note assuming maximal information transmission is consistent with the assumption of constant perception in the Weber-Fechner law, as only perceived gradients contain information for the cell. 

In the limiting cases that the input noise is much smaller or larger than the output noise, we can simplify this result. For small input noise, Eq.~\ref{eq:optimalDR} reduces to $\partial \bar{A}/\partial c \propto \sigma_A \cdot p_c(c)$, i.e. the input-output relationship is the steeper the higher the output noise \cite{TkaCal07a,TkaCal07b}. This can be understood intuitively, as different output levels become better separated (see Text S1). In the opposite case of large input noise, we obtain $\partial \bar{A}/\partial c \propto \sigma_A \sqrt[3]{ p_c(c)/\sigma_c^2}$. Hence, the larger the input noise the shallower the input-output relationship, as different input levels become better separated.
In the well-characterized chemotaxis system, where we have experimental dose-response curves determining the input-output relationship, we can thus predict the typical distributions of input concentrations using Eq.~\ref{eq:optimalDR}.

\subsection*{Predicting input and output distributions}

The distributions of input concentrations and cellular outputs can be predicted from experimental data based on the arguments presented above. Fig.~3A shows multiple dose-response measurements using FRET for fields of cells expressing only the Tar receptor.
As can be seen from the figure, there is large variability between different measurements of the curves. A large amount of this variability is caused by noisy receptor expression, most likely due to plasmid copy-number variation~\cite{EndSouWin2008}. 
Hence, in Fig.~3B (first column) we removed this expression noise from the variance using principal component analysis~\cite{EndSouWin2008} (see Text S1, although we later relax this assumption). 
The resulting standard deviation (SD) has a characteristic shape when plotted as a function of ligand concentration: cells adapted to zero background concentration (WT 2) and the QEEE mutant show a monotonically decreasing SD with increasing MeAsp concentration as receptors become saturated by attractant. Cells adapted to a higher ambient concentration (WT 2), as well as mutants in higher receptor-modification states, show a peak in the SD in the linear, steep regime of the dose-response curve.

To extract input ($\sigma_c^2$) and output ($\sigma_A^2$) noises for {\it single} cell we fitted the following intuitive equation for the variance of the total activity from $n \approx 400$ cells 
\begin{equation}
\sigma_{n,T}^2  = n^2 \underbrace{\alpha_1 c}_{\sigma^2_c} \left( \frac{\partial A}{\partial c} \right) ^2 + n \underbrace{\left[ \alpha_2 A \left(1-A\right)+\alpha_3 A \right]}_{\sigma^2_A},
\label{eq:deltaA2t}
\end{equation}
to the FRET variance $\text{SD}^2$ shown in the first column of Fig. 3B, with $\alpha_1$, $\alpha_2$ and $\alpha_3$ fitting parameters (listed in Table S1 with confidence intervals and $\chi^2$ quality-of-fit test), and $A$ the average activity of an individual cell from Eq.~\ref{eq:mwc} (a constant scaling factor between the receptor-complex and FRET activities is neglected here for simplicity). 
In Eq.~\ref{eq:deltaA2t}, the first, second, and third terms represent transmitted input noise, noise from switching of receptor complexes, and noise from the intracellular pathway, respectively. We assume all of these noise sources are independent. Specifically, input noise depends proportionally on concentration~$c$~\cite{BergPur77}, and is transmitted according to the mean input-output relationship given by the MWC model. Receptor switching depends on the probabilities of complexes to be \textit{on} and \textit{off}, and hence has a binomial activity dependence. Pathway fluctuations, e.g. due to dephosphorylation, are assumed to be Poissonian, and the number of phosphorylated signaling proteins is taken to be proportional to the signaling activity~$A$. 
In the first term the factor $n^2$ arises due to coherent addition of input fluctuations from all cells in the same flow chamber (e.g. due to pipetting errors). In the remaining noise terms, the factor $n$ arises due to the incoherent addition of cell-internal fluctuations from all cells.

Equipped with single-cell input-output relationships and noises we can predict the input and output distributions given optimal information transmission. Based on matching relation Eq.~\ref{eq:optimalDR}, the predicted distributions of concentration inputs are shown in Fig.~3B (second column).
For high receptor-modification states, the distributions can be fit by log-normal distributions as shown in Fig.~3C (fitting parameters are listed in Supporting Table S2 with confidence intervals). Log-normal distributions have the property that their ratio of variance and squared mean only depend on a single parameter. Hence, scaling log-normal distributions with equal relative variances such that their means coincide, collapses the entire functions on top of each other (inset of Fig.~3C). This constitutes further evidence for Weber's law and contrast coding. 
The quality of the scaling collapse is not further improved when removing all noise (equivalent to constant output noise), showing that noise is indeed small (or sufficiently uniform, see Fig. S4). 
Furthermore, scaling collapse and distributions of inputs do not change markedly when using total FRET variance for fit or using uniform output noise $\sigma_A^2$ in Eq. \ref{eq:deltaA2t} (see Fig. S4), demonstrating robustness of our predictions. 

The predicted output distributions (cf. Eq.~\ref{eq:optimalDR}) are shown in Fig.~3B (third column).  
Using the variance excluding the receptor-expression noise (first principal component), we obtain bimodal distributions for receptors in high modification states. This deviation from the flat maximum-entropy distribution (cf. Fig. S1B) is due to noise, and was observed in other biological systems \cite{TkaCal07a}.
Predicted mutual information values are $\sim 10$ bits, corresponding to a graded response with about $2^{10} \sim 1000$ distinguishable output levels. 
Such analogue computation is particularly useful when small variations in stimuli need to be sensed and processed \cite{Sarpeshkar_98}.

\subsection*{Distributions of chemical gradients from simulations of swimming bacteria}

Equipped with the distributions of sampled input concentrations, we aimed to characterize the chemical environment of swimming bacteria through the distributions of chemical gradients. For this purpose, we set up simulations of swimming bacteria using the RapidCell software \cite{VlaLovSou08} as described in {\it Materials and Methods}. Briefly, cells are able to sense and adapt via Eqs. 1-3, as well as to run and tumble based on cell-internal signaling. Cells are also subject to rotational diffusion.
To relate the distribution of gradients to the functional output (swimming behaviors) of chemotactic cells, we use the chemotactic index (CI), which measures the alignment of cell trajectories with the gradient often used to quantify eukaryotic chemotaxis \cite{Tweedy}, and the drift velocity ($v_d$) up the gradient \cite{ReneGopa10}. In fact, CI and drift velocity are closely related to each other (see Text S1). To achieve the above we follow a multistep protocol.

First, we validated our simulations by comparing trajectories of swimming bacteria with experimental tracking data in two dimensions in linear gradients~\cite{MassonCelaniVerg2012}. Specifically, we successfully compared the CI (Fig. S5A-D), and interval-length distributions of motor rotation (see Fig. S5E,F and \cite{BlockSegBerg83}). In addition to the tests conducted in \cite{VlaLovSou08}, this provides confidence that our simulations reproduce realistic swimming behavior.

Second, we set up simulations of adapting Tar-only bacteria swimming in different linear gradients along the $x$ direction. That way we obtained distributions of sampled concentrations at a particular receptor-modification level, denoted by $m^*$.
The idea is that in a very shallow gradient cells sample a symmetric range of concentrations around $c^*=c(m^*,A^*)$, with $A^*$ the adapted activity. However, when the gradient becomes steeper cells start moving up the gradient. As a result, cells sample a wider range of concentrations up the gradient than down the gradient, producing an asymmetric distribution of sampled concentrations (Fig. 4A). Since a dose-response curve has evolved to sense gradients best that cells most likely encounter, we compared the predicted distributions from information theory (cf. Fig. 3) to the sampled concentrations from simulations in order to select typical gradients.

As shown in Fig.~4B for one exemplar linear gradient, we extracted $100$s-long trajectories from simulations of about 1000 cells. From these trajectories we then calculated the CI as a function of concentration, and repeated this for different linear gradients. 
The resulting CI is a complicated, non-monotonic function of concentration and gradient, shaped by receptor sensitivity, adaptation, and rotational diffusion (see Fig. 4C, Text S1 and Fig. S6A,B for details).
In particular, for a given $c^*$ there are relative gradients, defined as gradient over $c^*$,  which lead to large CI values (red curve in Fig. 4C) while for both smaller (blue curve) and higher (green curve) relative gradients the CI values are reduced. In contrast, for smaller concentrations ($c<c^*$) a smaller gradient maximizes CI (blue curve), while for larger concentrations ($c>c^*$) a larger gradient maximizes CI (green curve). This indicates that chemotaxis is optimized with respect to the gradient for a given $m^*$ (and hence $c^*$).

Third, after obtaining the sampled concentrations from trajectories with $m=m^*$ (indicated by red lines in Fig.~4B), we used the overlap between these and the predicted input distributions to select the gradients sensed best. 
As shown in Fig.~5A for $m^*=4$ (see Figs. S7-S9 for more details and plots), only intermediate relative gradients give a sampled distribution that matches the prediction with significant overlap. In extremely shallow relative gradients the histograms of concentrations sampled at $m=m^*$ are symmetric due to adaptation, leading to low overlap and reduced CI values (Fig.~5A, left panel). On the other hand, extremely steep gradients lead to very asymmetric distributions. Although these cells locally move up the gradient more than predicted, their receptor signals are saturated and their long runs lead to low CI (Fig.~5A, right panel). 
In other words, to efficiently move up the gradient (maximum of CI) a cell has to balance between long runs up the gradient and sufficient tumbles to correct for misalignment.

In line with our expectation, Fig. 5B shows that maximal overlap between the simulated and predicted distributions of concentrations corresponds to maximal CI, and hence maximal drift up the gradient. 
Unlike CI, which encodes information on trajectories only, drift is about the speed in a gradient and thus can be considered the final output of the chemotactic cells. Drift is most likely the quantity under evolutionary pressure \cite{Skoge_Meir_Win_PRL11}. Also note in Fig. 5B that CI and drift fall off to zero at very shallow relative gradients as expected due to precise adaptation, but that they fall off more slowly for steep relative gradients. This is consistent with our prediction of a limiting CI value for steep gradients (see Text S1) and the strongly asymmetric distributions of sampled inputs in Fig. 5A (right). CI and drift become ill defined once the relative gradient reaches the inverse run length ($\sim 50$ $\text{mm}^{-1}$).

To finally extract the distributions of gradients, we realize that for exponential gradients, $c(x)=A e^{\lambda x}$,  predicted from the Weber-Fechner law the relative gradient is given by $(dc/dx)c^{-1}=\lambda=$ const, i.e. independent of $x$ and therefore $c$. As a consequence, if cells prefer exponential gradients with a particular rate value $\lambda$, then this preference should apply to all receptor-modification levels (as long as cells are responsive for $K_a^{\text{off}} \ll c \ll K_a^{\text{on}}$). Thus, when plotted as a function of relative gradient, we expect that all overlap, CI and drift curves collapse for different receptor-modification levels, which they approximately do (see Fig.~5B). The resulting distribution of exponential gradients is remarkably broad (wide range of rate values in Fig. 5D), demonstrating the wide dynamic range of the sensory system.  Consistently, the broadness index defined in {\it Materials and Methods} is markedly larger than one, i.e. ranges from about 3.7 for QEQQ to 5.5 for WT 2 (0.1 mM). Slightly broader distributions are obtained when fitting to the total variance from FRET (Fig. S10). This case may imitate strong gene-expression noise which sometimes even arises when expressing from the chromosome \cite{KollLovSou05, Spudich76}.


\section*{Discussion}

In the large intestine, hundreds of different bacterial species are present at high cell densities~\cite{PouLanKrog94}, forming a dynamic microenvironment with complex spatio-temporal chemical gradients from partially digested food and host secretions~\cite{MitKog06}.
Previous models accounted for Weber's law and fold-change detection (FCD) \cite{KalTuWu09,LazAhmShi2011,VlaLovSou08}, but were unable to make predictions beyond average gradients. Here, we first critically reevaluated FCD and Weber's law, also addressing the Weber-Fechner law. To predict distributions of concentration we then applied information theory to {\it in vivo} FRET data in {\it E. coli}. By comparing with simulations of swimming bacteria in different linear gradients, we found that the predicted distributions of input concentrations from information theory are consistent with bacteria experiencing a broad range of exponential gradients. In particular, fluctuating gradients (cell-external noise) and methylation levels (cell-internal noise) were needed to explain the long tails of the distributions in line with expectation \cite{ClarkGrant05,CelVerg2010}. Importantly, our results show that maximizing information transmission leads to maximizing the chemotactic index (CI), 
and hence the drift velocity up the gradient (Fig. 5) \cite{ReneGopa10}. Although expected, the existence of such a link between information transmission and drift has recently been questioned in bacterial chemotaxis \cite{Skoge_Meir_Win_PRL11}. Our model is thus able to connect a large number of phenomenological laws with information theory at the molecular level.

In contrast to the report by Lazova et al.~\cite{LazAhmShi2011}, we found that at large concentration changes the resulting large-amplitude responses and their subsequent adaptation curves do not show FCD (cf. Fig.~2C, D). A possible reason for this discrepancy is that Lazova {\it et al.} used small temporal variations of stimuli which varied on a time scale close to adaptation~\cite{LazAhmShi2011}. Hence, their experiments might have probed FCD with respect to small response amplitudes with temporal responses mainly determined by the external oscillatory stimulus rather than by internal signaling~\cite{BarLei97}.

The Weber-Fechner law requires a quantity called ``perception'', which has not been identified previously in bacterial chemotaxis. The free-energy difference between the {\it on} and {\it off} states of the receptor complex fulfills this purpose; it is a logarithmic function of the concentration in the validity range of Weber's law, thus leading to a compressed internal representation of cell-external stimuli.
This functional form arises from the molecular interaction between ligand and receptor, in particular from the competition between the gain of binding-free energy and loss of ligand-volume entropy upon ligand-receptor binding~\cite{KeyEndSko06}. As logarithmic response functions are implicated for ligand-receptor interactions \cite{VlaLovSou08, JiangOuyangTu10}, as well as simple membrane potentials (Nernst equation), our findings may apply to a wide range of signal transduction processes in cells~\cite{Lau89,KoshGoldStock82,KoesterInKan91}.

Information theory can predict the distributions of input concentrations, and using simulations the distributions of gradients. Interestingly, our predicted distribution of inputs are log-normal, similar to the identified scale invariance of light intensities in the visual system \cite{OlsFie00}. This observation may contain deeper insights into the workings of sensory systems. Specifically, log-normal distributions arise from multiplicative interactions in molecular components \cite{Frank13}. Indeed, our dose-response curves closely resemble Hill equations due to receptor cooperativity \cite{KeyEndSko06} (note Hill equations are the cumulative distributions of log-normal distributions). Our deduced exponential gradients of aspartate may arise from local nutrient sources and degradation by bacteria in the surroundings. 

Although our information-theoretic/simulation results are consistent with exponential gradients predicted by phenomenological laws, our model is based on a number of assumptions. These include small Gaussian noise and a certain functional form of the noise (Eq. \ref{eq:deltaA2t},  both assumptions were successfully tested in Fig. S4). Most importantly, we assume maximal information transmission. However, Tar receptors are also used for pH and temperature sensing \cite{Tu14PLoSCB}, and global optimization may result in suboptimal sensing of individual stimuli types \cite{Lander13}. Additionally, expression of chemotactic genes is linked to nutrient supply, growth conditions, and life cycle \cite{McCarter06}, further 
restricting the predictive power of our information-theoretic approach. 
Furthermore, this approach only applies to (nearly) instantaneous receptor signaling, and hence excludes slower downstream dynamics, in particular of the motor~\cite{TosttenWolde09}. 
However, we expect that if information transmission is optimized by the whole chemotaxis pathway, this should also be true for every intermediate stage, as information can only be lost, not gained. Note that our data of non-adapting cells was restricted to Tar-only cells. While this restricted the dynamic range, Tar is the cognate receptor for our stimuli.

%
To fully characterize the microenvironment of {\it E. coli} future work may need to factor in additional types of stimuli, such as other chemicals, temperature \cite{OleksiukWinSou11} and pH \cite{HuTu13}. Models can first be tested in well-defined gradients of stimuli produced in microfabricated devices~\cite{HasPost07,MassonCelaniVerg2012,HuTu13,WolPil08,KangHanLee}, thus establishing the stimuli cells sense best. 
By measuring FRET in single cells \cite{Vaknin_Berg_04}, the fidelity of information transmission can be tested more directly without any masking effects at the population level. This may also help answering whether information is conserved between graded receptor signaling and the binary-like motor response. Ultimately, imaging and tracking bacteria in complex microenvironments will enhance our understanding of chemotaxis under natural conditions.

\section*{Materials and Methods}

\subsection*{FRET measurements}

FRET measurements were performed as described previously \cite{SouBerg02a,OleksiukWinSou11}, using phosphorylation-dependent interaction between the response regulator CheY, fused to yellow fluorescent protein (CheY-YFP), and its phosphatase CheZ, fused to cyan fluorescent protein (CheZ-CFP), as a readout of the pathway activity.

\subsection*{Parameters of MWC model}

For the MWC model for receptor complexes we use the following parameters as derived from fits to FRET data~\cite{KeyEndSko06,ClauOleEnd10}: $K_a^\text{off}=0.02\,\text{mM}$, $K_a^\text{on}=0.5\,\text{mM}$, $K_s^\text{off}=100\,\text{mM}$ and $K_s^\text{on}=10^6\,\text{mM}$. The size of receptor complexes is assumed to increase as a function of ambient concentration $c_0$ according to $N(c_0) = a_0 + a_1 c_0$, with $a_0$=17.5 and $a_1$=3.35/mM \cite{ClauOleEnd10}. Furthermore, we used a direct interpolation of data from \cite{EndSouWin2008} to evaluate $\epsilon (m)$ instead of using the equation $\epsilon (m) \simeq 1-0.5 m$, see Supporting Fig. S3 for the comparison.  
For the adaptation rate constants we used $g_R$=0.0069~s$^{-1}$ and $g_B$=0.11~s$^{-1}$~\cite{ClauOleEnd10}.
For the concentration-step profiles we assumed exponential rising and falling functions with rate constants $\lambda_\text{add}$=0.6~s$^{-1}$ and $\lambda_\text{rem}$=0.5~s$^{-1}$~\cite{ClauOleEnd10,SouBerg02a}.

\subsection*{Maximizing information transmission with input and output noise}

We assume a Gaussian channel, i.e. the input-output relationship is given by a dose-response curve for the mean $\bar{A}(c)$ and normally distributed output noise with total variance $\sigma^2_T = (\partial \bar{A} / \partial c)^2 \sigma^2_c + \sigma^2_A $, consisting of transmitted cell-external (input) and cell-internal (output) noise, respectively.
The conditional probability for the output given the input is then
\begin{equation}
 p(A|c) = \frac{1}{\sqrt{2\pi\sigma_T^2(c)}}\exp\left\lbrace-\frac{[A-\bar{A}(c)]^2}{2\sigma^2_T(c)}\right\rbrace.
\end{equation}
The mutual information is given by Eq.~\ref{eq:mutualinformation_gaussianchannel}~\cite{DetRamShr00}, which is maximized with respect to $\bar{A}$ using the Lagrange formalism. Introducing gain $G = \partial \bar{A} / \partial c$ and interpreting the integrand of Eq.~\ref{eq:mutualinformation_gaussianchannel} as the Lagrangian
\begin{eqnarray}
 {\cal{L}} \!\!\! & = \!\!\! & p_c(c)\log\left[\sqrt{ \sigma^2_A G^{-2} + \sigma^2_c } p_c(c)\right] \nonumber\\
 \!\!\! &= \!\!\!  & p_c(c) \left[ \frac{1}{2} \log( \sigma^2_A + \sigma^2_c  G^2) + \log(p_c(c)) - \log(G) \right],
\end{eqnarray}
the following Euler-Lagrange equation is obtained
\begin{equation}
 \frac{\partial {\cal{L}}}{\partial \bar{A}} - \frac{d}{dc}\frac{\partial {\cal{L}}}{\partial G} = 0.
\end{equation}
The derivatives of ${\cal{L}}$ with respect to $\bar{A}$ and $G$, respectively, evaluate to
\begin{eqnarray}
 \frac{\partial {\cal{L}}}{\partial \bar{A}} \!\!\!& =\!\!\! & \frac{1}{2} \frac{p_c(c) \frac{\partial \sigma^2_A}{\partial \bar{A} }}{ \sigma^2_A  + \sigma^2_c G^{2}} \\
 \frac{\partial {\cal{L}}}{\partial G} \!\!\!& =\!\!\! & -\frac{p_c(c)  \sigma^2_A}{ G (\sigma^2_A  + \sigma^2_c G^{2}) }.
\end{eqnarray}
The derivative of the last term with respect to the input variable $c$ is
\begin{equation}
\frac{d}{dc}\frac{\partial {\cal{L}}}{\partial G} = - \frac{ \frac{\partial p_c(c) }{\partial c}  \sigma^2_A }{ G( \sigma^2_A + \sigma^2_c  G^2 )}   -  p_c(c) \frac{d}{dc} \left[ \frac{ \sigma^2_A }{ G(\sigma^2_A+ \sigma^2_c  G^2)}\right].
\end{equation}
The resulting Euler-Lagrange can be rearranged into
\begin{equation}
-\frac{\frac{d\xi}{dc}}{\xi} = \frac{G}{2 \sigma^2_A} \frac{\partial \sigma^2_A}{\partial \bar{A}} + \frac{\frac{\partial p_c(c) }{\partial c} }{p_c(c)}\label{eq:resEulerLag}
\end{equation}
with $\xi = \sigma^2_A / G (\sigma^2_A  + \sigma^2_c G^{2})$. Integration over the input $c$ yields
Eq.~\ref{eq:optimalDR} for the dose-response curve $G= \partial \bar{A} / \partial c$.

\subsection*{Experimental tracking and simulations of swimming cells in spatial gradients}

Experimental data for bacteria in linear gradients were obtained from Jean-Baptiste Masson~\cite{MassonCelaniVerg2012}. To generate the dynamics of the receptor activity and modification level, we used the information about sampled concentrations at each position as input for the MWC model. We initialized cells with the adapted modification level corresponding to the initial concentration and integrated Eqs.~\ref{eq:mwc}-\ref{eq:preciseadaptation}. 

To simulate swimming Tar-only cells, we used the software package RapidCell~\cite{VlaLovSou08}. The modification dynamics within RapidCell were replaced by our pathway model with parameters as described in the text. This led to an adapted activity $A^*\approx 1/3$. The simulation box was set to a rectangle of 5~mm in $y$- and variable dimension in $x$-direction, allowing bacteria with 5 flagella each to sample concentrations from 0-0.8 mM (note that beyond 0.6 mM cells stop responding and only run \cite{EndWin06}). Periodic boundary conditions were applied. Cells reaching the boundary were taken out and injected at the opposing side. The receptor-modification level was then set to the adapted level corresponding to the concentration at the injection site  with tumbling to reinitialize the swimming direction. For analysis we excluded a 0.2~mm border both for comparison with the data by Masson et al.~\cite{MassonCelaniVerg2012} and for the prediction of gradient distributions.
For the prediction of gradients, linear gradients in $x$-direction with different slopes and base concentration were used for identification of the range of well-sensed gradients and calculation of the chemotactic index (CI). Randomly shifting the base concentration by normal distribution with standard deviation of 0.001\%  allowed simulation of input noise. 
To calculate histograms of sampled concentrations, all cells with modification level equal to $m^*$ of the corresponding strain were selected using a normal distribution with relative standard deviation 0.04\% (thus mimicking output noise in line with \cite{ClauEnd11BMC}). These noise sources broadened the distributions of sampled inputs to better match the predicted distribution of inputs. 

For the prediction of gradients, the overlap between the obtained histograms $p(c,m^*)$ and the predicted input distributions ($p_{\text{input}}$) were calculated using $\int \text{min}\left(p_{\text{input}}; p(c,m^*)\right) \mathrm d c$ using software R (version 3.0.1). The calculation of CI and drift is detailed in the Text S1. The broadness of the overlap (broadness index) was estimated by the ratio of fold change in overlap to fold change in relative gradient between their respective minimal and maximal values. The latter are defined by achieving an overlap within 20\% of maximal overlap.


\section*{Acknowledgments}

We thank Nikita Vladimirov for providing the RapidCell software~\cite{VlaLovSou08}, Jean-Baptiste Masson and Massimo Vergassola for sharing the trajectories from microfluidics, Frank J$\ddot{\mathrm{u}}$licher for helpful discussions, as well as Nick S. Jones and Karen Ottemann for constructive comments on the manuscript. 

\bibliography{all}


\begin{figure}[!ht]
\begin{center}
\includegraphics[width=\textwidth]{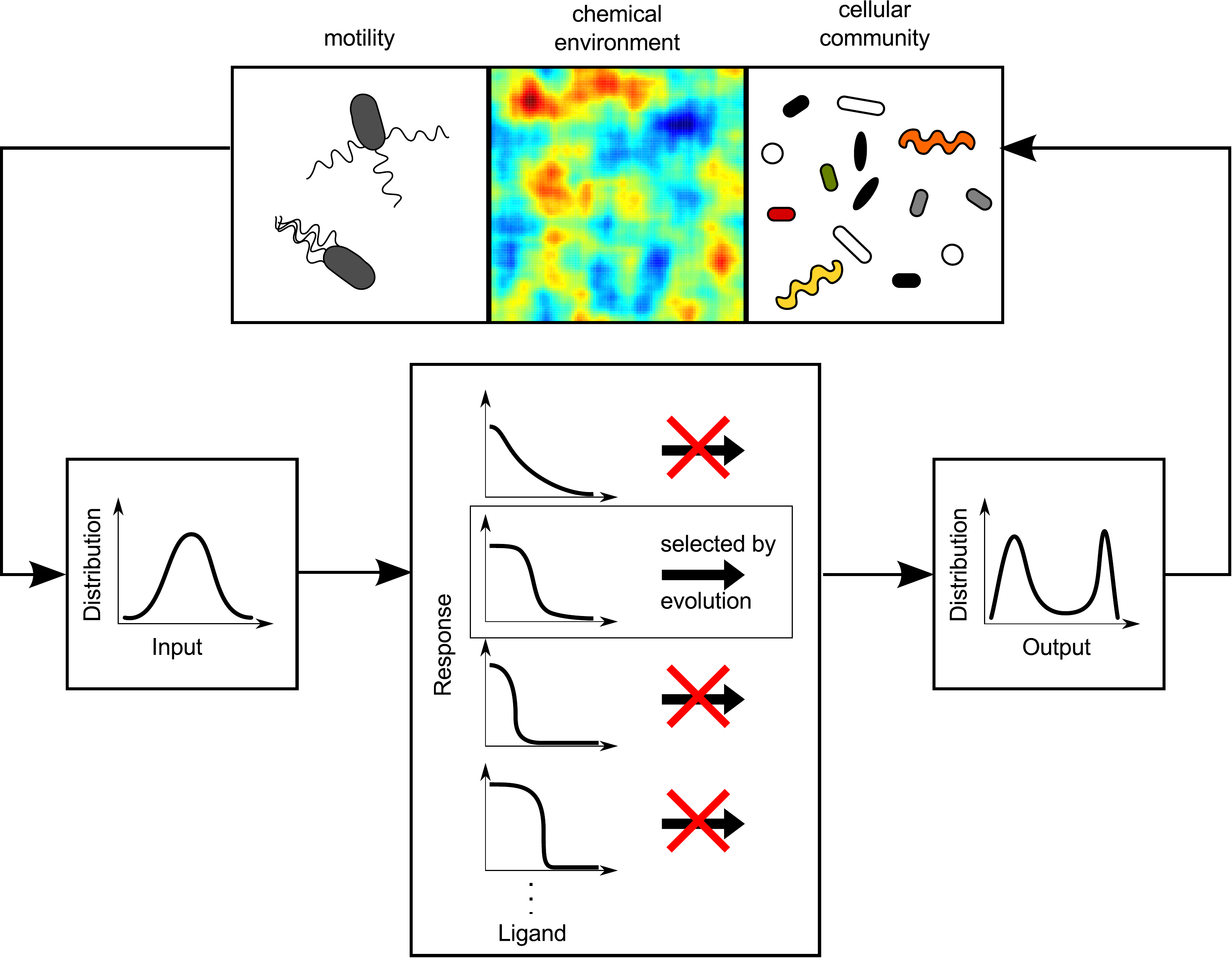}
\end{center}
\caption{
{\bf The {\it Escherichia coli} chemosensory system is adapted to its chemical environment by evolution.} Influences such as motile behavior, chemical sources (e.g. patchy food gradients in the human intestine), and the multitude of other organisms shape the typical concentrations sampled by a bacterium, leading to typical input distributions of chemical concentrations. Through signal transduction the sensory system produces (intracellular) output distributions. Evolution is expected to have selected the optimal shape of the input-output (dose-response) curve to allow for an appropriate response to typical stimuli.
}
\label{fig:fig1}
\end{figure}

\begin{figure}[!ht]
\begin{center}
\includegraphics[width=0.625\textwidth]{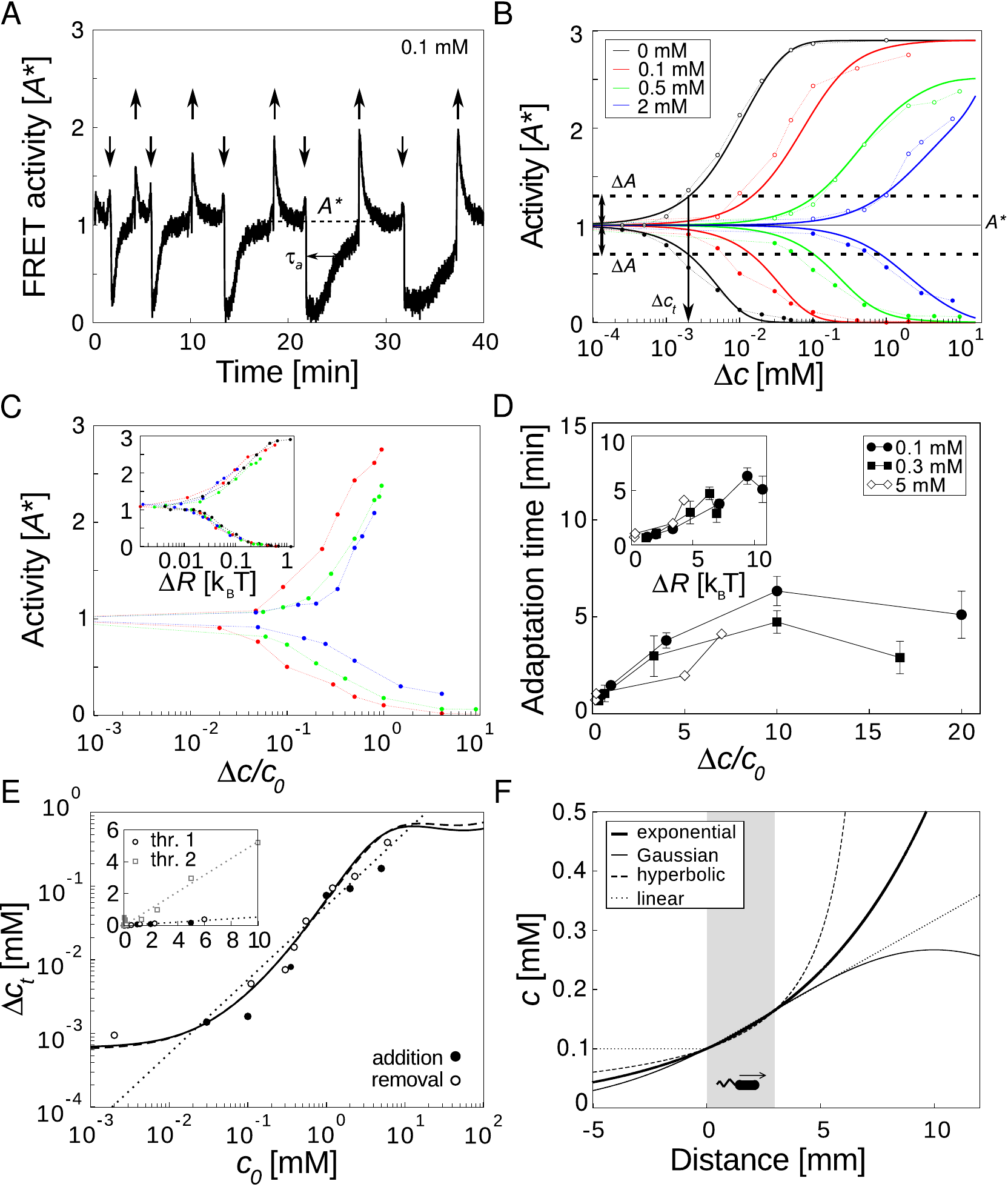}
\end{center}
\caption{
{\bf Weber's law and fold-change detection of adapting wild-type (WT 1) cells.} (A)~Example time course of a FRET measurement in response to added ($\downarrow$) and subsequently removed ($\uparrow$) step changes of MeAsp (background concentration 0.1 mM MeAsp). The adapted activity $A^*$ and the adaptation half-time $\tau_a$, the time to reach half of the pre-stimulus activity are indicated. (B)~Dose-response curves for cells adapted to ambient concentrations of 0, 0.1, 0.5 and 2 mM MeAsp and subjected to increasing concentration step changes $\Delta c$ of MeAsp (curves from left to right). Filled and open circles correspond to response to addition and removal of attractant, respectively~\cite{SouBerg02a}. Solid lines represent the MWC model of mixed Tar/Tsr-receptor complexes~\cite{ClauOleEnd10}. FRET and MWC model activities were normalized by adapted pre-stimulus values at each ambient concentration. The adapted activity $A^*$ (thin black line) and example threshold activities $A^*\pm \Delta A$ (dashed-
dotted lines) for removal (upper curve) and addition (lower curve) of concentration step changes are also indicated. The threshold stimulus $\Delta c_t$ corresponds to the concentration change $\Delta c$ where the activity reaches the threshold activity. (C) Dose-response curves from panel~B plotted as function of fractional changes in concentrations $\Delta c/c_0$ for background concentrations of 0.1, 0.5 and 2 mM MeAsp. ({\it C, Inset}) Dose-response curves from panel~B plotted as function of changes in perception $\Delta F$ for background concentrations 0, 0.1, 0.5 and 2 mM MeAsp. Note 0 mM background curve can be plotted as a function of $\Delta R$ but not as a function of $\Delta c/c_0$. (D) Adaptation half-times for various concentration step changes as function of the fractional concentration change for 0.1, 0.3 and 5 mM MeAsp ambient concentration. ({\it D Inset}) The same adaptation times as a function of change in perception. (E)~Threshold stimulus $\Delta c_t$ for addition and removal of MeAsp to 
achieve the response threshold $\Delta A\!\!=\!\!0.08 A^*$, plotted as function of background concentration $c_0$ (data, symbols). MWC model (solid and dashed lines) and a linear fit of the experimental data points (dotted line) are shown as well. ({\it E, Inset}) Experimental data points (symbols) and linear fits (dotted lines) are plotted on a linear scale for two different response thresholds. Threshold 1: $\Delta A\!\!=\!\!0.08$ $A^*$ (circles), threshold 2: $\Delta A\!\!=\!\!0.16$ $A^*$ (squares). Filled and open symbols represent addition and removal of MeAsp, respectively. Slopes of linear fits are 0.054 (threshold 1) and 0.115 (threshold 2). (F)~Different spatial concentration profiles with equal concentrations at distances $x$=0 and $x$=3 mm, spanned by grey box. 
}
\label{fig:fig2}
\end{figure}

\begin{figure}[!ht]
\begin{center}
 \includegraphics[width=\textwidth]{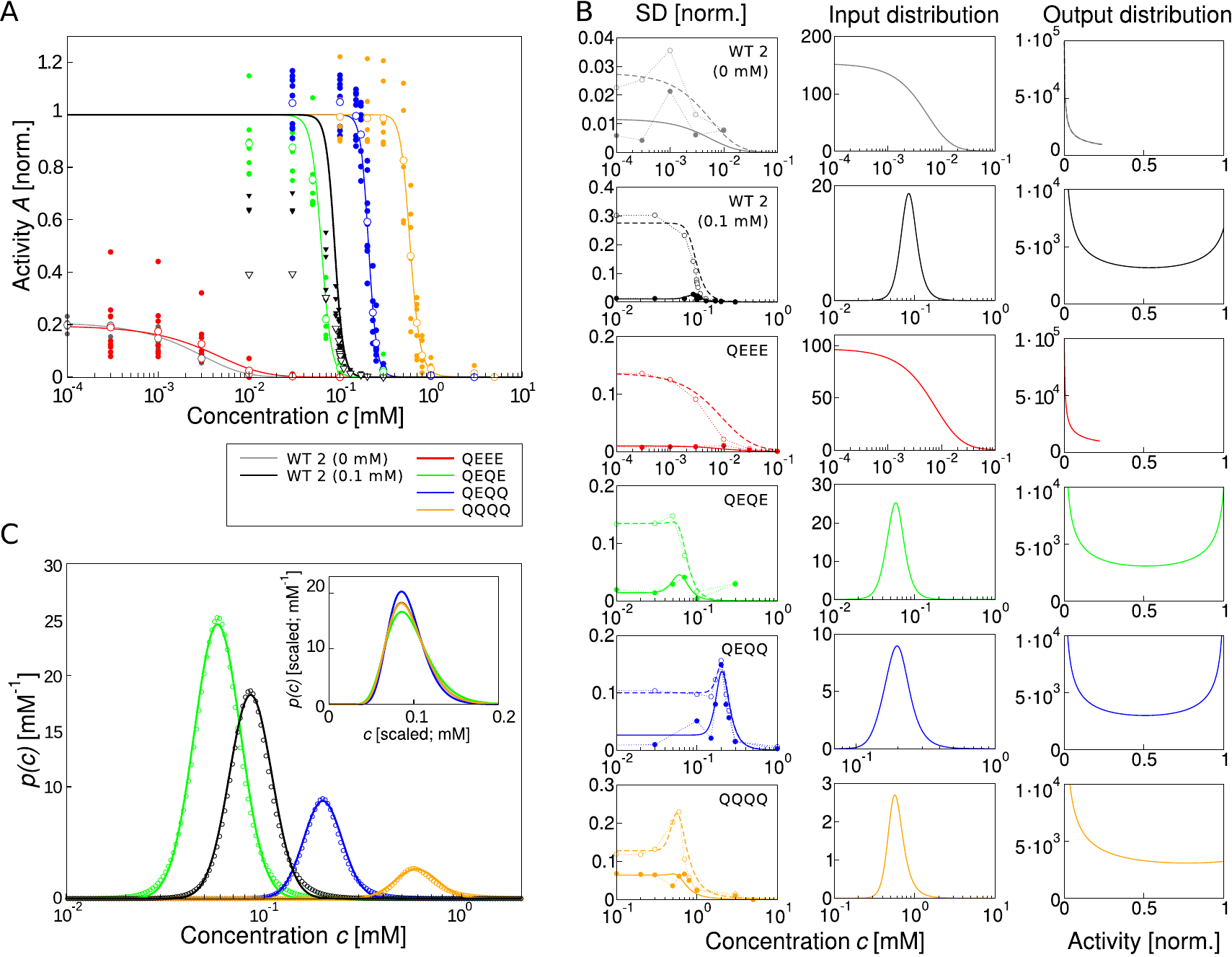}
\end{center}
\caption{
{\bf Predicted distributions of chemical inputs and intracellular outputs.}  (A)~Dose-response curves for adapting (WT 2) cells and non-adapting mutants in specific receptor-modification states as taken from~\cite{EndSouWin2008}. Cells express the Tar receptor using an inducible plasmid with all receptors removed from chromosome. Data points for various measurements of the FRET activity at different concentrations of MeAsp (filled symbols), as well as their mean values are shown (open symbols). Strains: cells adapted to zero 0 mM MeAsp(black circles) and 0.1~mM MeAsp (black triangles) background concentration, QEEE (red), QEQE (green), QEQQ (blue) and QQQQ (orange) mutant. Corresponding solid lines are the fits of the MWC model~\cite{EndSouWin2008}. Receptor complex and FRET activities were normalized to maximal activity. 
(B)~Predicted input and output distributions. ({\it Left}) Standard deviation (SD) of activity measured by FRET in panel~A (open and filled symbols for including and excluding receptor expression noise, respectively). Fits to the standard deviations are shown as well (dashed and solid lines, respectively). ({\it Middle}) Predicted distributions of attractant concentrations (input). Note input distributions for low activity curves (0 mM MeAps and QEEE) look noticeably different from high activity curves (0.1 mM MeAsp, QEQE, QEQQ, and QQQQ). ({\it Right}) Predicted distributions of signaling activities (output).
(C)~Log-normal scaling of the predicted input concentrations. All curves are normalized with area of one. Symbols are the predicted input distributions panel~B (QEQE, green; WT 2 (0.1~mM), black; QEQQ, blue; QQQQ, orange) and colored lines are log-normal fits to the distributions. ({\it Inset}) Log-normal fits from the main panel scaled to the peak position of the input distribution from cells adapted to 0.1 mM MeAps.
See Tables S1 and S2 for fitting parameters with assessment of confidence.
The calculated mutual information between inputs and outputs is $\sim$10 bits for different modifications from adapting and mutant cells (WT, 23.3; WT 0.1mM, 7.7; QEEE, 6.6; QEQE, 6.8; QEQQ, 4.5; QQQQ, 18.9 in bits).
}
\label{fig:fig3}
\end{figure}

\begin{figure}[!ht]
\begin{center}
\includegraphics[width=8.3cm]{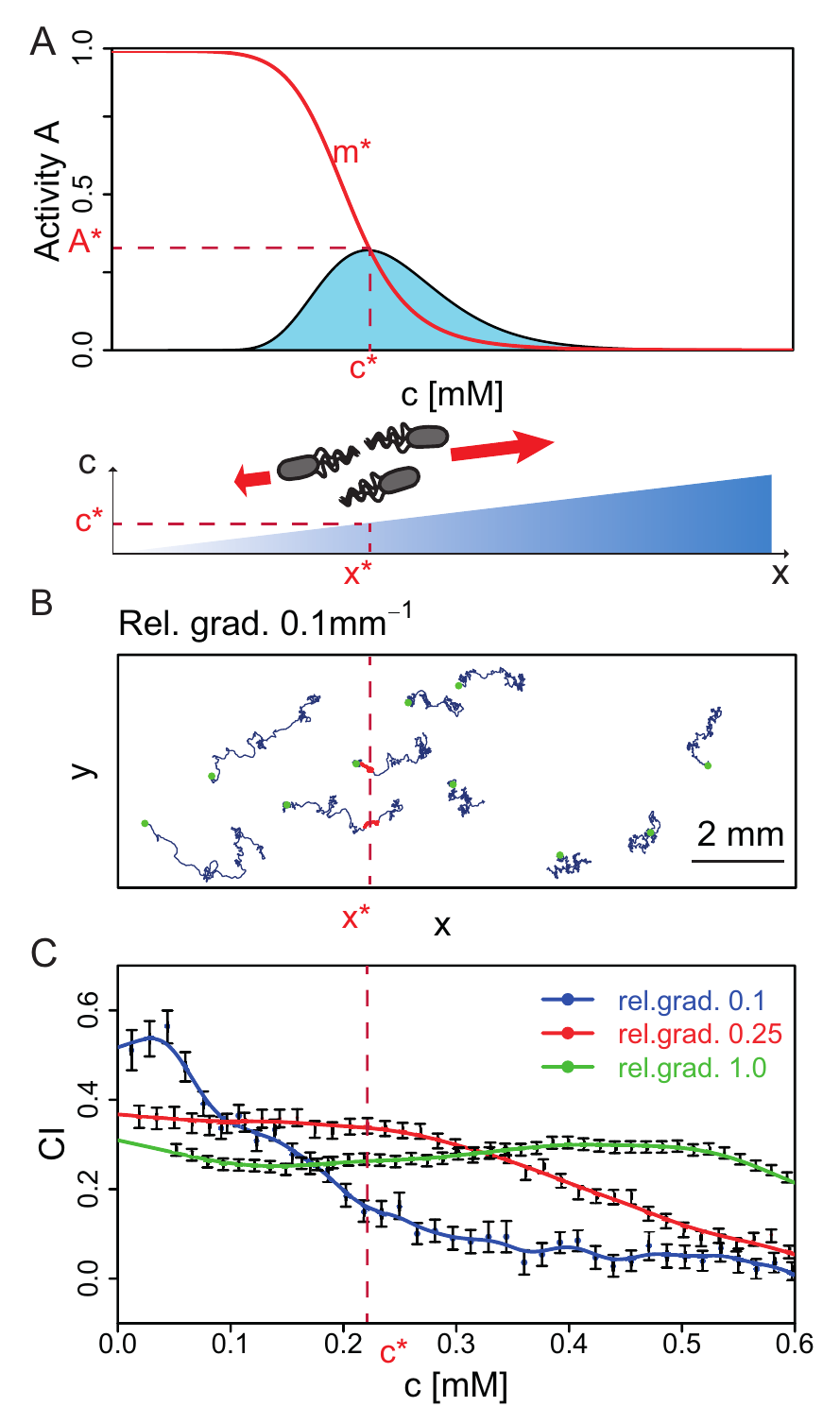}
\end{center}
\caption{
{\bf Simulations of swimming bacteria in chemical gradients.}  (A)~Strategy to extract sampled distribution of inputs from simulations. Adapted activity $A^*$ and modification level $m^*$ determine concentration $c^*$ (and position $x^*$). Cells swimming in typical gradients should sample the same concentrations as predicted by information theory (light blue curve). (B)~Simulation in a rectangular box with gradient in $x$ as described in {\it Materials and Methods}. Trajectories of Tar-only cells for relative gradient $0.1\text{mm}^{-1}$, using $m^*=6$ (QEQQ) as an example. Green dots, blue lines and red lines correspond to initial positions, full trajectories, and partial trajectories with modification level of bacteria equal to $m^*$, respectively. (C)~Chemotactic index (CI) extracted from all simulated trajectories in the box for different relative gradients as function of concentration in the box. Gradients relative to $c^*=223.9 \mu$M: 0.1$\text{mm}^{-1}$ (blue), 0.25$\text{mm}^{-1}$ (red) and 1.0$\text{
mm}^{-1}$ (green). Non-monotonic trend of CI as function of $c$ suggests an optimal range of gradients which maximizes CI. For example, at $c^*$ the maximum CI corresponds to linear relative gradient 0.25~$\text{mm}^{-1}$. Symbols and error bars indicate averages and standard errors of the mean from several trajectories at this concentration and gradient. Lines are interpolations by smooth functions. }
\label{fig:fig4}
\end{figure}

\begin{figure}[!ht]
\begin{center}
\includegraphics[width=\textwidth]{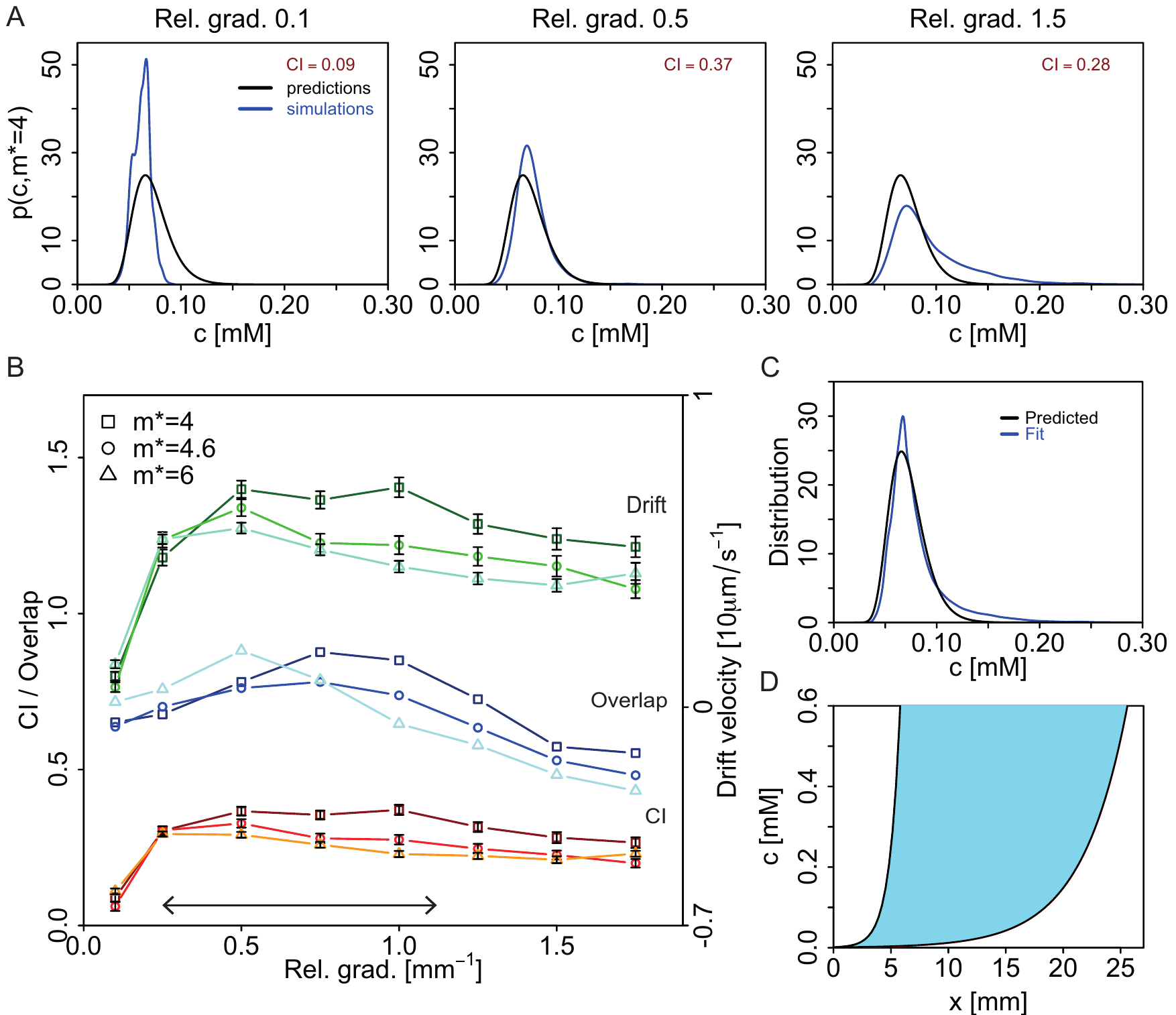}
\end{center}
\caption{
{\bf Reconstruction of distributions of sampled gradients.}  (A)~Using $m^*=4$ (QEQE) as an example, predicted distribution of inputs from information theory (black lines) and distributions of sampled concentrations (blue lines) obtained for cells swimming in increasing relative linear gradients, $0.1 \text{mm}^{-1}$ (left), $0.5 \text{mm}^{-1}$ (central), and $1.5 \text{mm}^{-1}$ (right) (gradients relative to $c^*=0.065$mM). To imitate cell-external noise, the base concentration of the gradients was fluctuating every 0.1$s$ with standard deviation~0.001 mM. To imitate cell-internal noise, modification level was selected from normal distribution with relative standard deviation $\sigma=0.04$ in line with previous results \cite{ClauEnd11BMC}. 
(B)~Overlap between distribution of sampled concentrations and predicted distribution (blue shades), chemotactic index (CI, red shades) and drift velocity (green shades) with modification level $m=m^*$. Symbols indicate modification level: squares, circles and triangles stand for $m^*=4$ (QEQE, $c^*\simeq0.07$ mM), $m^*=4.6$ (WT 2, $c^*\simeq0.09$ mM), and $m^*=6$ (QEQQ, $c^*\simeq0.22$ mM), respectively. $m^*=8$ (QQQQ, $c^*=0.63$ mM) is not included as Tar-only cells do not adapt at high values of $c^*$. Horizontal arrow illustrates range of relative gradients over which the overlap is within 20\% of maximal value on average between the three modification levels. 
(C)~Sampled distributions from different relative gradients (0.1-1.75$\text{mm}^{-1}$) indeed fit prediction (overlap 90\%). 
(D)~Range of exponential gradients predicted to be sensed best (blue area), according to the range indicated by horizontal arrow in (B).
}
\label{Figure_label}
\end{figure}

\end{document}